\documentclass[preprint]{aastex631}

\usepackage{mathptmx}
\usepackage[T1]{fontenc}
\usepackage{ae,aecompl}
\usepackage{graphicx}
\usepackage{amsmath}
\usepackage{amssymb}
\usepackage{subfigmat}
\usepackage{enumerate}
\usepackage{multirow}
\usepackage{natbib}

\usepackage{tabularx, booktabs}
\usepackage{rotating}
\usepackage{longtable}
\usepackage{ltablex}
\usepackage{threeparttablex}

\def\gtrsim{\mathrel{\hbox{\rlap{\hbox{\lower5pt\hbox{$\sim$}}}\hbox{$>$}}}}

\newcolumntype{Y}{>{\centering\arraybackslash}X}

\shorttitle{PCA filter}
\shortauthors{Yun \& Lee}
\accepted{2023 September 14}
\submitjournal{\apj}

\begin{document}

\correspondingauthor{Jeong-Eun Lee}
\email{lee.jeongeun@snu.ac.kr}

\title{The PCA Filtering method for an unbiased spectral survey of Complex Organic Molecules (COMs)}

\author[0000-0001-6842-1555]{Hyeong-Sik Yun}
\affiliation{Korea Astronomy and Space Science Institute, 776 Daedeok-daero, Yuseong, Daejeon 34055, Korea, hsyun@kasi.re.kr}

\author[0000-0003-3119-2087]{Jeong-Eun Lee}
\affiliation{Department of Physics and Astronomy, Seoul National University, 1 Gwanak-ro, Gwanak-gu, Seoul 08826, Korea, lee.jeongeun@snu.ac.kr}
\affil{SNU Astronomy Research Center, Seoul National University, 1 Gwanak-ro, Gwanak-gu, Seoul 08826, Republic of Korea}

\begin{abstract}
A variety of interstellar complex organic molecules (COMs) have been detected in various physical conditions. However, in the protostellar and protoplanetary environments, their complex kinematics make line profiles blend each other and the line strength of weak lines weaker. In this paper, we utilize the principal component analysis (PCA) technique to develop a filtering method which can extract COM spectra from the main kinematic component associated with COM emission and increase the signal-to-noise ratio (SNR) of spectra. This filtering method corrects non-Gaussian line profiles caused by the kinematics. For this development, we adopt the ALMA BAND 6 spectral survey data of V883 Ori, an eruptive young star with a Keplerian disk. A filter was, first, created using 34 strong and well-isolated COM lines and then applied to the entire spectral range of the dataset. The first principal component (PC1) describes the most common emission structure of the selected lines, which is confined within the water sublimation radius ($\sim$ 0.3 \arcsec) in the Keplerian disk of V883 Ori. Using this PC1 filter, we extracted high-SNR kinematics-corrected spectra of V883 Ori over the entire spectral coverage of $\sim$50 GHz. The PC1-filtering method reduces the noise by a factor of $\sim$ 2 compared to the average spectra over the COM emission region. One important advantage of this PC1-filtering method over the previously developed matched filtering method is to preserve the original integrated intensities of COM lines. 
\end{abstract}

\section{Introduction}
Molecules that contain six or more atoms, including carbon, are called complex organic molecules \citep[COMs; ][]{Her09}. These molecules are also the building blocks of life. Recently, interstellar COMs have been discovered in various physical and dynamical conditions \citep[]{Cec22}. Particularly, COMs are found in the inner regions of protostellar envelopes and protoplanetary disks, where the icy mantle on dust sublimates above the temperature of $\sim$ 100 K \citep{JELee19,Bia22,Tob23}. Detection of various COMs near protostars raises intriguing questions about how and when these COMs would be delivered to forming planets. To answer these questions, we need to investigate, with unbiased spectral surveys, the COM evolution along the star formation process from envelope to disk.

Thanks to the unprecedented high sensitivity and high resolution provided by the Atacama Large Millimeter/submillimeter Array (ALMA), a variety of COMs recently have been detected in various interstellar environments, such as infrared dark clouds \citep{Sak18}, low- and high-mass protostars \citep{Bog19,Cse19,Man20,Yan21}, and protoplanetary disks \citep{Fav18,CFLee19,JELee19}. The ALMA spectra of protostellar inner hot envelopes, so-called hot cores or hot corinos, show a forest of COM lines \citep{Jor16,Jor18,Sak18,Man20,Bel20,Hsu22}, which are often blended together due to kinematically broadened line profiles. In addition, weak lines become more diluted due to the kinematically broadened line shape and consequently are embedded within the noise \citep{Yen16,Loo18}.

\citet{JELee19} extracted COM spectra from V883 Ori by aligning the centroid velocities and integrating the emission at different positions \citep{Yen16}. The authors measured the centroid velocities using C$^{17}$O $J=$3$-$2, which is strong enough to derive velocities accurately. The spatial distribution of COM emission, however, differs from that of the C$^{17}$O emission in V883 Ori \citep{JELee19}.

\citet{Loo18} introduced a matched filtering method to obtain high signal-to-noise ratio (SNR) spectral lines from interferometric data sets. The matched filtering method derives a filter response spectrum using an expected emission distribution in Fourier space as a kernel. This technique is an efficient way to perform rapid line identification without an imaging process \citep{Loo18,Boo19,Loo20}. However, for further quantitative analysis, the specific lines still must be imaged.

Both the methods, by \citet{JELee19} and by \citet{Loo18}, require prior information on the spatial and spectral distribution of line emission. However, since different molecules often trace different gas components \citep{Tyc21}, the prior information obtained via strong lines from simple molecules may not directly correspond to the spatial and spectral distribution of the COM lines. With ALMA, unbiased spectral survey observations of COMs have been carried out \citep[e.g.,][]{Jor16}. For efficient analyses of the 1000 to 10,000 COM lines covered by the spectral survey observations toward a given target, we require a method to extract COM spectra showing the correct associated kinematics while preserving the original integrated intensities without any prior information.

Principal component analysis (PCA) is one of the multivariate statistics that access the common features veiled in the variation of data. The PCA has been applied to spectral cube data of line emission at radio wavelengths, uncovering not only the properties of the turbulent environments \citep{Hey97,Bru13,Yun21} but also the chemical variation \citep{Ung97,Lo09,Jon13,Gra17} within molecular clouds. For example, \citet{Oko21} applied PCA to ALMA observations toward L483 to assess the distribution of molecular lines in position-position-velocity spaces (PCA-3D). The authors found that the first-order principal component (PC1), which describes a typical distribution of molecular lines \citep{Ung97,Lo09,Jon13,Gra17}, showed a representative velocity structure of the rotating disk/envelope of L483.

V883 Ori is an outbursting FU Orionis-type object in transition from the Class I to Class II stages. Previous ALMA observations revealed a well-developed rotating protoplanetary disk around the central source \citep{Rui17,vant18}. For the first time, many COM lines, including CH$_3$OH, CH$_3$CHO, CH$_3$COCH$_3$, CH$_3$OCHO, and CH$_3$CN \citep{JELee19}, were detected from the disk. The COM lines are detected within a small area within r $\sim$ 120~au (0.3\arcsec). \citet{Tob23} detected water (HDO and H$_2^{18}$O) lines and estimated the water snowline, which is the water sublimation radius in the disk midplane, at $\sim$ 80~au in V883 Ori. Because of the temperature structure, the water sublimation front develops the 2-dimensional surface in the disk and extends to $\sim$0.3\arcsec\ in the disk surface of V883 Ori \citep{JELee19, Tob23}. In this paper, the water sublimation radius refers to the water sublimation radius in the disk surface.

Recently, Lee et al. (submitted) observed V883 Ori with the ALMA spectral scan mode in Band 6 covering $\sim$55~GHz via the project {\it The ALMA Spectral Survey of An eruptive Young star, V883 Ori (ASSAY)}. Various simple molecules and COMs have been detected, and it has been demonstrated that different molecules trace different kinematical components within V883 Ori. Among the detected lines, as presented in \citet{Tob23}, the HDO line is detected toward the inner region of the protoplanetary disk within the projected radius of $\sim$120 au (0.3\arcsec), and the COMs emission is confined within the water sublimation region. To identify all the detected COM lines and derive the physical conditions of the gas component associated with COM emission, high-SNR line spectra corrected for the kinematics are essential. Similar to \citet{Oko21}, a typical emission distribution of COM lines can be derived by applying PCA to a sample of COM lines, which are well-isolated and relatively strong. By utilizing this common emission distribution, we can collect emission signals extended through the position-position-velocity space and make a single Gaussian-like emission line for a specific transition.

This paper introduces a new filtering method to extract the high-SNR kinematics-corrected spectra of COMs by applying PCA to the unbiased ALMA spectral survey data of V883 Ori (ASSAY). Our filtering method aims to obtain high-SNR spectra with kinematics-corrected Gaussian-like line profiles, still preserving their original intensities. Section \ref{sec_obs} describes the ALMA Band 6 data of V883 Ori. The methodology of the filtering method is explained in Section \ref{Sec_anal}. We will assess how the filtering process can improve the spectra in Section \ref{sec_rst}. Section \ref{sec_disc} discusses the intensity preservation of the filtered spectra and their applications. Finally, we summarize our results in Section \ref{sec_sum}.

\section{Data} \label{sec_obs}
We adopt the Cycle 7 ALMA Band 6 observation of the project ASSAY (ALMA Spectral Survey of An eruptive Young star, V883 Ori; 2019.1.00377.S, PI: Jeong-Eun Lee) from Lee et al. (submitted). These ALMA data were obtained for an unbiased spectral survey of V883 Ori covering from 220.7 to 274.4~GHz. The full data set comprises three Science Goals (SGs), and each SG contains 20 spectral windows (SPWs), resulting in 60 SPWs in total. The velocity resolution of image cubes varies from 0.662 (at 221~GHz) to 0.533~km~s$^{-1}$ (at 274~GHz). All images are convolved such that their final beam properties match the poorest beam, whose size is 0.25\arcsec\ $\times$ 0.15\arcsec\ with position angle -77$\degr$. The rms noise temperature ($T_\mathrm{rms}$) varies from 1.143 to 1.915~K with a mean value of 1.481~K. Refer to Lee et al. (submitted) for further details of the observations and data reduction. 

To evaluate the spectra produced by our filtering method, we compare them with three sets of spectra, (1) aperture-averaged spectra, (2) aligned spectra, and (3) spectra extracted using the matched filtering method. A typical method to obtain a line spectrum is deriving a mean spectrum from a given aperture that covers the line emitting region. Since the COM emissions are detected within the water sublimation region, we obtain the {\it aperture-averaged spectra} using a circular aperture with a radius of 0.3\arcsec\ from each spectral cube data. 

We also generate line spectra using the same method as \citet{JELee19}. This method aligns the line central velocity at each pixel, which is shifted by the disk rotation of V883 Ori, to the source velocity of 4.3~km~s$^{-1}$, and extracts an average spectrum within an aperture \citep{Yen16}. During this procedure, we determine the line central velocities at each pixel by utilizing the intensity-weighted velocity (moment 1) map of C$^{17}$O $J=$2$-$1 at 224.7144~GHz. This line is one of the strongest lines and is well known to trace a rotating disk (\citealt{vant20}; Lee et al. submitted). We extract average spectra for each SPW using an aperture covering the water-sublimated region (r $\leqq$ 0.3 \arcsec) on a de-projected image of V883 Ori. These spectra are referred to as the {\it aligned spectra}. We also obtained additional aligned spectra from an narrow annulus at the radial distance from 0.145\arcsec\ to 0.155\arcsec\, specifically isolating strong COM lines. These aligned spectra from the narrow annulus is adopted for initial line inspection (see Section \ref{Sec_line_sel}). 

The matched filtering method is applied to the cube data for the second SPW of the first SG, which covers the frequency range from 221.63 to 222.54~GHz. To obtain the {\it matched-filtered spectrum}, we created a kernel for matched filter spanning $\pm$ 10~km~s$^{-1}$ relative to the system velocity within 0.3\arcsec\ from the center of V883 Ori using the C$^{17}$O line.

Measuring the 1-$\sigma$ noise level is important to determine the sensitivity of the filtered spectra. However, the error propagation for the filtering method is challenging since many spatial and spectral pixels are considered together to accumulate the line emission. We thus derive noise spectra for each SPW by applying filtering to emission-free image cubes extracted from non-primary-beam-corrected images. These emission-free cubes are referred to as the {\it reference images}. To obtain the reference images, we selected a 1\arcsec\ $\times$ 1\arcsec\ area located at 5\arcsec\ east from V883 Ori.

The rms noise levels of the aperture-averaged and aligned spectra are also measured using these reference images. Assuming that the reference images are centered on V883 Ori, we obtained the emission-free aligned and aperture-averaged spectra using the same approach as described above. Subsequently, we then measured the rms noise level of the aligned and aperture-averaged spectra from their corresponding emission-free spectra. Note that we cannot compute an emission-free matched-filtered spectrum since matched filter method is applied to visibility data. Thus the rms noise level of the matched-filtered spectrum is derived from the emission-free channels within the second SPW of the first SG.

\section{Analysis} \label{Sec_anal}
\subsection{Line selection} \label{Sec_line_sel}
It is important to select appropriate lines for the PCA. We first identified detected molecules using the eXtended CASA Line Analysis Software Suite \citep[XCLASS; ][]{Mol17}. 71 molecules, including isotopologues and isomers, have been identified by matching against $\sim$3,000 detected lines in the spectral survey data. The detailed procedures of this analysis will be described in a separate paper. Next, we investigated the line transitions of the detected molecules from the Cologne Database for Molecular Spectroscopy \citep[CDMS; ][]{Mul01,Mul05} database and the Jet Propulsion Laboratory \citep[JPL; ][]{Pic98} catalog, where 61,541 lines exist in total within the covered frequency range. 

Among these transition lines, we selected specific lines based on several criteria. These criteria include the lines being detected above the 5-$\sigma$ noise level in the aligned spectra extracted from the narrow annulus, having the upper state energies (E$_\mathrm{up}$) lower than 2000~K, Einstein coefficient A (A$_\mathrm{ij}$) greater than 10$^{-8}$, and being isolated from the other candidate lines. We define a line as an isolated line if it does not have any other lines within $\pm$5~km~s$^{-1}$ of its line center. The 5~km~s$^{-1}$ criterion for isolation was determined through iterative line selection processes, taking into account that the full width of the zero intensity of the selected lines is slightly smaller than 10~km~s$^{-1}$. As a result, We identified a total of 34 {\it isolated strong COM lines} (see Appendix \ref{App_isol_str}).

\subsection{Computing the PCA}
The PCA-3D is applied to all 34 isolated strong lines to obtain the representative velocity and emission structure of the COMs within V883 Ori. We construct a data set for the selected lines, which share the same spatial and velocity spaces. For each of the selected lines, its cube data is extracted from a 1\arcsec\ by 1\arcsec\ area centered on V883 Ori, within which most of the COM emission arises (the white dotted box in the right bottom panel in Figure \ref{fig_IS_lines_mom0}). This limited image size prevents the introduction of noise-dominated principal components (PCs) in the PCA by excluding noise spikes from emission-free areas. All the extracted cube data cover the velocity space of $\pm$ 5~km~s$^{-1}$ with respect to the system velocity of 4.3~km~s$^{-1}$, while they have slightly different velocity resolutions ($\Delta v$) depending on the line frequencies. To better identify the velocity structure traced by the COM emission, we interpolate the extracted cube data with the finest $\Delta v$ of 0.533~km~s$^{-1}$.

The PCA adopted in this study is the same as that used by \citet{Oko21}. A correlation matrix is used to derive the Principal Components (PCs) of the 34 lines. This method avoids the PCs being dominated by a single line. We obtain the correlation matrix ($c_{ij}$) of the lines as follows,
\begin{equation}
  c_{ij} = \mathrm{corr}(T_i,T_j); T_i = T(X_k,v_m)
\end{equation}
where corr($T_i$,$T_j$) is the correlation coefficient between $i$th and $j$th lines, and $T_i$ is the intensity of $i$th line from a position of $X_k$ at a velocity $v_m$. The 34 PCs (34 eigenvectors and corresponding eigenvalues) can be obtained by diagonalizing $c_{ij}$. For each PC, an eigenvector ($u_n$) contains component scores for each of the 34 lines. The component scores describe how the intensities of the 34 lines correlate. The eigenvalue ($\lambda_n$) represents the portion of total variation explained with the corresponding PC. The PCs are ordered by decreasing $\lambda_n$; the first PC (PC1; n = 1) has the largest eigenvalue ($\lambda_1$), explaining the largest portion of the total variation.

An eigencube ($T_{\mathrm{PC}n}$), a dot product of line datacubes ($T_i$) and component scores for a given order $n$, reveals the intensity distribution explained by a corresponding PC$n$;
\begin{equation}
  T_{\mathrm{PC}n}(X_k,v_m) = \sum_{i = 0}^{n_\mathrm{line}} T_i(X_k,v_m)u_n(i),
  \label{eqn_2}
\end{equation}
where $n_\mathrm{line}$ is the number of the analyzed lines and $u_n$($i$) is the component score for the $i$th line. We produce the eigencube of PC1 to obtain the representative velocity structure of the 34 isolated strong lines \citep{Oko21}.

\subsection{PC1 Filtering method}
One of the essential benefits of the PCA filtering process is accumulating line emission in different velocities and spaces to one Gaussian line profile centered at the system velocity. This is possible because $T_\mathrm{PC1}$($X_k$,$v_m$) describes how the COM lines distribute over the spatial ($X_k$) and velocity spaces ($v_m$). For example, in a Keplerian disk, the line profile at a given position typically has a Gaussian shape due to thermal and non-thermal broadening. However, the central velocity of the Gaussian profile is shifted by the disk rotation. Therefore, the line profile extracted over the entire disk has a double-peaked line shape \citep{Smak1981}. However, if we know how much the central velocity shifts at individual positions, then we can correct the velocity shifts to the source velocity before accumulating line emission over the disk to produce a single Gaussian profile. It is, in principle, the same process implemented to stack spectra using prior information on the velocity structure of V883 Ori \citep{JELee19}. 

 We derive a window function for the filtering method, $f$($X_k$,$v_m$) by normalizing $T_\mathrm{PC1}$($X_k$,$v_m$):
\begin{equation}
  f(X_k,v_m) = \frac{T_\mathrm{PC1}(X_k,
v_m)}{\sum_{X_k} \sum_{v_m = v_\mathrm{min}}^{v_\mathrm{max}} T_\mathrm{PC1}(X_k,v_m) \Delta v}, \label{eqn_window}
\end{equation}
where $v_\mathrm{min}$ and $v_\mathrm{max}$ are the minimum and maximum velocities of the window function with ($v_\mathrm{sys}-5$)~km~s$^{-1}$ and ($v_\mathrm{sys}+5$)~km~s$^{-1}$, respectively. Since the window function is normalized, the PC1-filtering method can preserve the filtered line intensities, another essential part of this filtering method. Subsequently, the filtered spectrum ($T_\mathrm{filt}$) is derived as follows,
\begin{equation}
  T_\mathrm{filt}(v_m) = \sum_{X_k} f(X_k,v_m)*T_\mathrm{obs}(X_k,v_m).
\end{equation}
Here $f*T_\mathrm{obs}$ denotes a convolution of the window function and the observed ALMA Band 6 cube data ($T_\mathrm{obs}$),
\begin{equation}
  f(X_k,v_m)*T_\mathrm{obs}(X_k,v_m) = \int_{v_\mathrm{min}}^{v_\mathrm{max}} f(X_k,v')T_\mathrm{obs}(X_k,v_m-v') dv'.
\end{equation}

\section{Result} \label{sec_rst}
\subsection{Results of the PCA} \label{sec_rst_PCA}

\begin{figure*}[htp]
\centering
\includegraphics[width=0.9\textwidth]{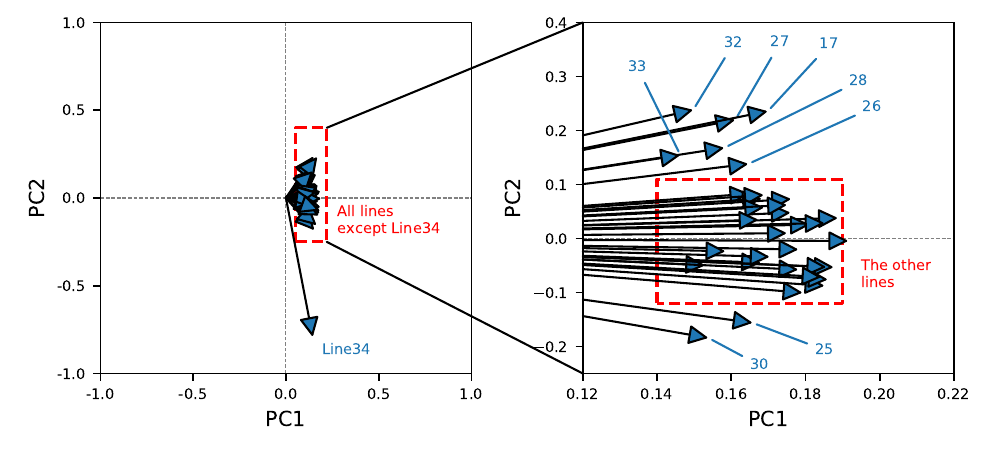}
\caption{A correlation wheel plot for the PC1 and PC2. The right panel shows a zoomed-in-view for detail and the numbers in blue refer to the COM number identified in Table \ref{tbl_isol_str}.}
\label{fig_PC12_corr_wheel}
\end{figure*}

The derived $\lambda_1$ is 23.73, and the fraction of the total variation of the data described by PC1, $P_\mathrm{1}$ = 69.8\%. Here
\begin{equation}
  P_\mathrm{1} = 100 \frac{\lambda_1}{\sum_{n=1}^{n_\mathrm{line}} \lambda_n}.\\
\end{equation}
Otherwise, the other PCs (from PC2 to PC34) have relatively small values of $\lambda_n$ (from 0.09 to 0.57). Therefore, the majority of the total variation is described by PC1, while the remaining 30.2\% of the total variation is described almost evenly by the other 33 PCs. Figure \ref{fig_PC12_corr_wheel} shows a correlation wheel plot for PC1 and PC2. The correlation wheel plot is a plot of arrows defined by a pair of component scores from a pair of PCs, and it is a popular way to present a variation of the emission lines described by PCs \citep{Ung97,Lo09,Pet17}. All arrows in the correlation wheel point to the right side: all component scores of PC1 have a positive sign varying from 0.14 to 0.19. It implies that all 34 lines positively correlate with each other. As a result, PC1 describes most of the total variation of 34 isolated strong lines, and they have similar emission distributions across the spatial and velocity spaces.

\begin{figure}[htp]
\centering
\includegraphics[width=0.4\textwidth]{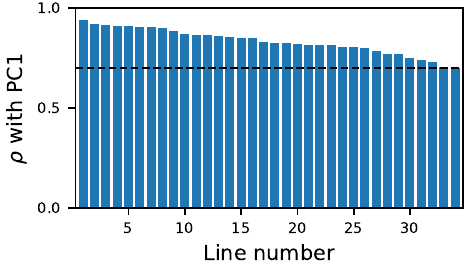}
\caption{Pearson correlation coefficients ($\rho$) between the eigencube of PC1 and the spectral cube data for each isolated strong line. The black dashed line indicates $\rho$ = 0.7.}
\label{fig_PC1_corr}
\end{figure}

We also calculate the Pearson correlation coefficients between $T_\mathrm{PC1}$ and $T_\mathrm{obs}$ for 34 COM lines to assess the distribution of lines \citep{Oko21}. Figure \ref{fig_PC1_corr} shows the correlation coefficients, which are ordered by their values (from Line1 to Line34; Table \ref{tbl_isol_str}). The correlation coefficients for all lines are larger than 0.7 (the black dashed line). This result is consistent with the result from the correlation wheel plot: all lines are tightly correlated with $T_\mathrm{PC1}$, confirming that $T_\mathrm{PC1}$ provides a common distribution of the 34 isolated strong lines. 

\begin{figure*}[htp]
\centering
\includegraphics[width=0.9\textwidth]{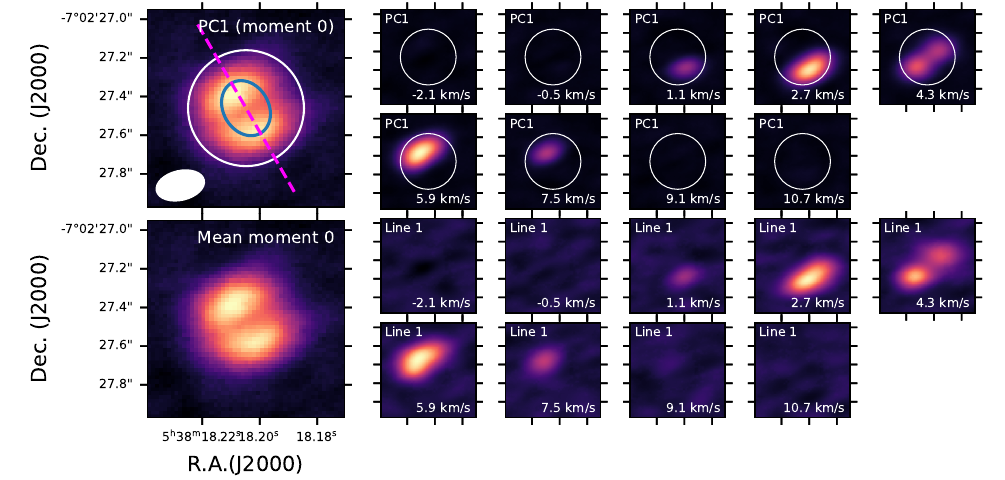}
\caption{The moment 0 and channel maps of the eigencube of PC1 ($T_\mathrm{PC1}$) (upper) and observed lines (lower). In the upper panels, the white circles mark a radius of 0.3\arcsec, within which most of the emission features arise. 
On the moment 0 map of PC1, the white-filled ellipse at the left bottom represents the beam shape, and the magenta dashed line indicates a cut for the position-velocity diagram, Figure \ref{fig_PV}, along the semi-major axis of the protoplanetary disk. The blue ellipse represents the radial distance of 0.15\arcsec, along which the aligned spectra for the line selection were extracted, on the de-projected image of V883 Ori. The lower-left panel shows the mean moment 0 map of the isolated strong lines, while the lower-right small panels present the channel maps for the CH$_3$OH line at 261.8057~GHz (Line 1; see Table \ref{tbl_isol_str}).}
\label{fig_PC1_Eigcube}
\end{figure*}

Figure \ref{fig_PC1_Eigcube} exhibits the moment 0 (upper-left panel) and channel maps (upper-right panels) of $T_\mathrm{PC1}$. The moment 0 map of PC1 resembles the mean moment 0 map for the isolated strong lines (lower-left panel). Most line emission arises from a small circular area with a radius of 0.3\arcsec\ (the solid white line). Two bright blobs are located at the north-eastern and south-western parts of the target, and a narrow region between the two bright blobs has a dip in line intensity. \citet{Loo18} and \citet{vant20} also showed a similar emission distribution for protoplanetary disks. The high-resolution images show a crescent shape of emission distribution with a central emission hole within r$\sim$0.1\arcsec\ \citep{JELee19, Tob23}, but the elongated beam shape of SG3 distorts the emission distribution in the images used in this study (Lee et al. submitted). In addition, the central emission hole, which is produced by the optically thick dust emission, is also distorted by the elongated beam shape, resulting in the dip between two elongated emission blobs. The channel maps show the emission features expected from a disk with a Keplerian rotation. The position-velocity (PV) diagram along the disk's semi-major axis exhibits emission features consistent with a Keplerian rotation profile around a 1.2~M$_\sun$ central protostar (see Appendix \ref{App_PV}).

\subsection{PC1-Filtering of spectra}
\subsubsection{A test with the PC1 itself} \label{sec_test}
Before applying the PC1-filtering method to the entire spectral survey data, we check how the filtering method corrects the complex line profile of the observed data. For this, we first adopt the $T_\mathrm{PC1}$ itself as test data. Applying the PC1-filtering method to $T_\mathrm{PC1}$ can validate the correction of the observed line profiles, as it tightly correlates with the 34 isolated strong lines. For comparison, we extract an average spectrum of $T_\mathrm{PC1}$ in the same way that is used to obtain the aperture-averaged spectra (see Section \ref{sec_obs}). The intensity scaling of $T_\mathrm{PC1}$ is set by Equation \ref{eqn_2} and does not refer to any particular line strength. We thus normalize the average spectrum to have a peak intensity of one. The aperture-averaged spectrum of PC1 has a double-peaked line profile expected from a rotating disk.

\begin{figure}[htp]
\centering
\includegraphics[width=0.4\textwidth]{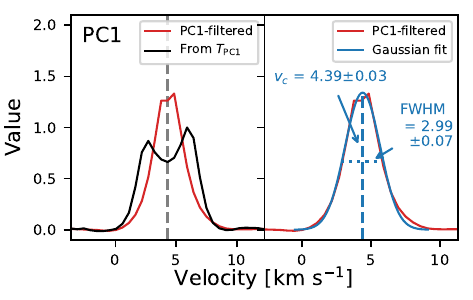}
\caption{Left: Correction of the line profile of PC1 using the PC1-filtering method. The black solid line represents a spectrum extracted from the PC1 eigencube with an aperture of r$=$0.3\arcsec. Note that the y-axis is normalized by the maximum value of the average spectrum. The PC1-filtered spectrum is presented as the red solid line. The gray dashed vertical line denotes the system velocity of 4.3~km~s$^{-1}$. Right: the PC1-filtered spectrum and its Gaussian fitting result (blue). The blue dashed and dotted lines represent the central velocity ($v_c$) and full-width half maximum (FWHM) of the Gaussian profile, respectively.}
\label{fig_filter_test}
\end{figure}

The left panel of Figure \ref{fig_filter_test} shows the PC1-filtered (red) and the average spectra (black). The filtering method changes the double-peaked line profile into a single-peaked profile. The peak intensity increases after the filtering. The filtered line profile is fitted by a Gaussian profile (blue) with a central velocity ($v_c$) of 4.39$\pm$0.03~km~s$^{-1}$ and an FWHM of 2.99$\pm$0.07~km~s$^{-1}$ (the right panel of Figure \ref{fig_filter_test}). This result proves that the PC1-filtering method can correct the complex line profile using the typical velocity and emission structure extracted by the PCA analysis.

\subsubsection{The PC1-filtering of the observed spectra} \label{sec_filt_obs}

\begin{figure*}[htp]
\centering
\includegraphics[width=0.9\textwidth]{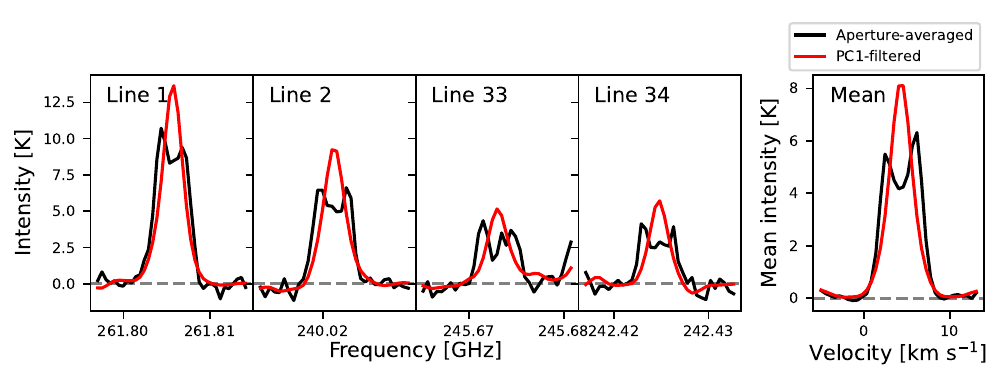}
\caption{Line profile corrections for the isolated strong lines. The four left panels show the aperture-averaged (black) and PC1-filtered (red) spectra for Lines 1, 2, 33, and 34. Lines 1 and 2 have the strongest correlation with the PC1, while Lines 33 and 34 have the weakest correlations. The right panel shows the mean spectra of 34 isolated strong lines.}
\label{fig_line_profile}
\end{figure*}

Now, we apply the PC1-filtering method to the real ALMA Band 6 data of V883 Ori. First, we investigate the changes in line profiles of 34 isolated strong lines. Figure \ref{fig_line_profile} shows examples of corrected line profiles by the PC1-filtering method; the double-peaked profile is corrected into a single-peaked profile with a higher peak intensity, as presented in Figure \ref{fig_filter_test}. Additionally, the filtered line profiles can also be described by a single Gaussian function; the mean filtered spectrum (the red spectrum in the right panel) can be fitted with a Gaussian profile with $v_c$ of 4.30$\pm$0.02~km~s$^{-1}$ and FWHM of 3.28$\pm$0.05~km~s$^{-1}$. This FWHM is slightly broader than the PC1 line profile corrected with the PC1-filtering method (see Figure \ref{fig_filter_test}). This discrepancy arises because $T_\mathrm{PC1}$ is filtered using itself (see Equation \ref{eqn_window}), resulting in an exact match between the window function and test data, i.e., $T_\mathrm{PC1}$. The convolution maximizes the peak intensity and leads to a slightly narrower line than the PC1-filtered observed lines.

\begin{figure*}[htp]
\centering
\includegraphics[width=0.9\textwidth]{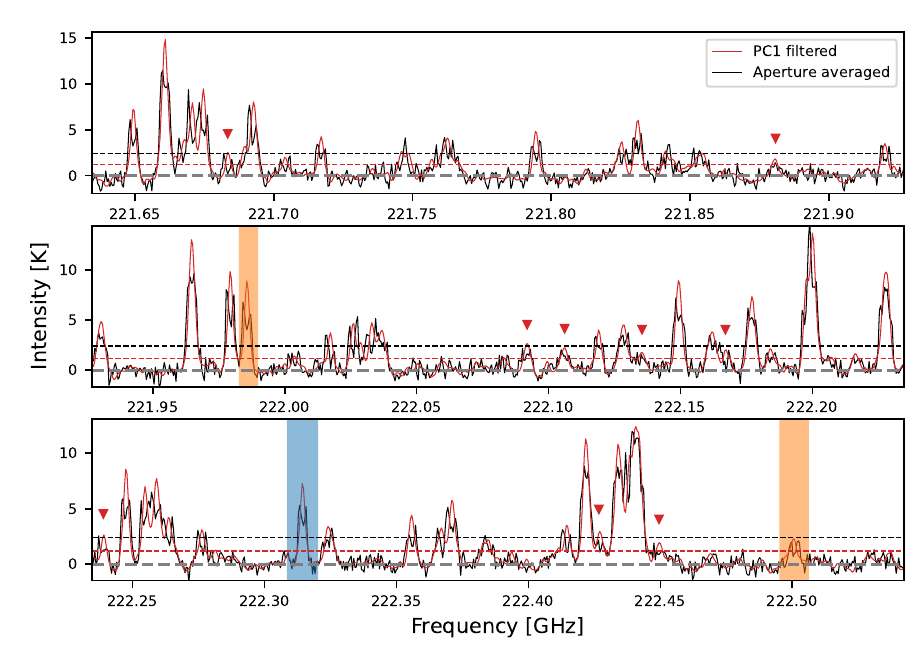}
\caption{The spectra for the second SPW of the first SG (from 221.63 to 222.54~GHz). The spectra are divided into three panels for better visibility. The aperture-averaged and PC1-filtered spectra are presented in black and red, respectively. The black and red horizontal dashed lines represent 3-$\sigma$ uncertainties in the aperture-averaged and PC1-filtered spectra, respectively. The blue shade indicates a member of the 34 isolated strong lines (see Table \ref{tbl_isol_str}), and the orange shades indicate the spectral ranges selected for checking conservation of total intensity (see Section \ref{sec_conserv}). The red triangles indicate some of the additional emission lines identified above the 3-$\sigma$ level in the PC1-filtered spectrum.}
\label{fig_filter_spec}
\end{figure*}

Finally, we apply the PC1-filtering method to the spectra obtained by the unbiased spectral survey of V883 Ori. Figure \ref{fig_filter_spec} presents the result only for the second SPWs of the first SG as an example. As anticipated, the filtered spectra (red) exhibit single-peaked Gaussian-like line profiles at the frequencies where the aperture-averaged spectra (black) present double-peaked line profiles. Also, the peak intensities of the observed lines in the PC1-filtered spectra are higher than those in the aperture-averaged spectra. This comparison is the same as in Figure \ref{fig_line_profile}, except for the PC1-filtering of the continuous spectral data over a wide frequency range rather than a specific individual line profile.

One notable feature of the filtered spectra is the improved SNR. The red and black dashed lines in Figure \ref{fig_filter_spec} indicate 3-$\sigma$ noise levels of the filtered and aperture-averaged spectra, respectively. The noise level of the PC1-filtered spectrum is measured using the reference image; the root mean squares calculated from the emission-free PC1-filtered spectra are adopted as the 1-$\sigma$ noise levels. The 1-$\sigma$ noise levels of the aperture-averaged spectra vary from 0.467 to 0.973~K, with a mean value of 0.694~K. On the other hand, those of the filtered spectra vary from 0.215 to 0.526~K, with a mean value of 0.329~K. The mean noise level decreases by a factor of $\sim$2 via the PC1-filtering method.

The noise level decreases by this filtering process because (1) noise is proportional to $\frac{1}{\sqrt{n_\mathrm{pix}}}$, where $n_\mathrm{pix}$ is the number of the spatial and spectral pixels included in the filtering process, (2) correlated noise signals in interferometric data are decorrelated by collecting emission lines in different velocities \citep{Yen16}, and (3) the contribution of the noise signal is reduced by low weight in the window function \citep{Loo18}. The SNR should increase by more than a factor of 2 since the peak intensity increases by the correction of velocity shift. Indeed, in this SPW, the SNR is enhanced by a factor of $\sim$ 2.5 on average, with the maximum improvement reaching around a factor of 3.0 for a transition line at 222.2474~GHz. The improved SNR in the PC1-filtered spectra enables the detection of additional COM lines. The red triangles in Figure \ref{fig_filter_spec} indicate some isolated transition lines. These lines are not detected in the aperture-averaged spectra, but are identified in the PC1-filtered spectra above the 3-$\sigma$ noise level. In total, we find 31 additional COM lines from the PC1-filtered spectra, which are $\sim$ 36 percentage of the detected lines in this SPW. This result demonstrates the effectiveness of our filtering method in discovering more COM lines from the observed data.

\section{Discussion} \label{sec_disc}
\subsection{Conservation of the total intensity.} \label{sec_conserv}

\begin{figure*}[htp]
\centering
\includegraphics[width=0.45\textwidth]{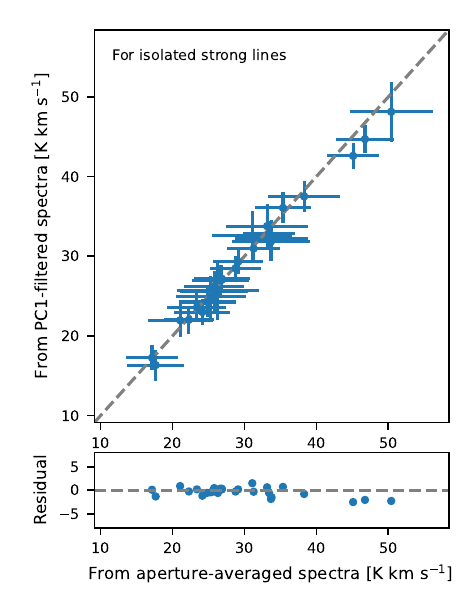}
\includegraphics[width=0.45\textwidth]{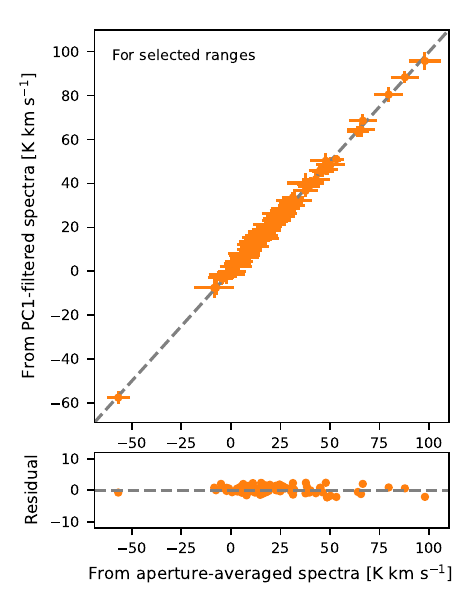}
\caption{The integrated intensities for 34 isolated strong lines (left) and 109 selected frequency ranges (right). The top panels show a comparison between the integrated intensities measured from the aperture-averaged and PC1-filtered spectra. The bottom panels show the difference in integrated intensity (residuals). The error bars in both panels represent the 3-$\sigma$ uncertainty of the measured integrated intensities.}
\label{fig_Int_conserv}
\end{figure*}

In this section, we compare the integrated intensities of individual lines to check whether the PC1-filtering method preserves line intensities. We measure the integrated intensities of 34 isolated strong lines from both the PC1-filtered and aperture-averaged spectra (see the left panel of Figure \ref{fig_Int_conserv}). The integrated intensity of each line is derived within $\pm$ 5~km~s$^{-1}$ from the line center. 

The integrated intensities from the PC1-filtered spectra tend to be slightly lower than those from the aperture-averaged spectra when the integrated intensity exceeds 40~K~km~s$^{-1}$. However, the difference in integrated intensity is still within a 3-$\sigma$ error range. The rms difference in integrated intensity of the isolated strong lines is about 1.07~K~km~s$^{-1}$. Since the mean error of the integrated intensities from the PC1-filtered spectra is about 0.7~K~km~s$^{-1}$, the integrated intensities from the PC1-filtered spectra are still consistent with those from the averaged spectra.

\begin{figure*}[htp]
\centering
\includegraphics[width=0.9\textwidth]{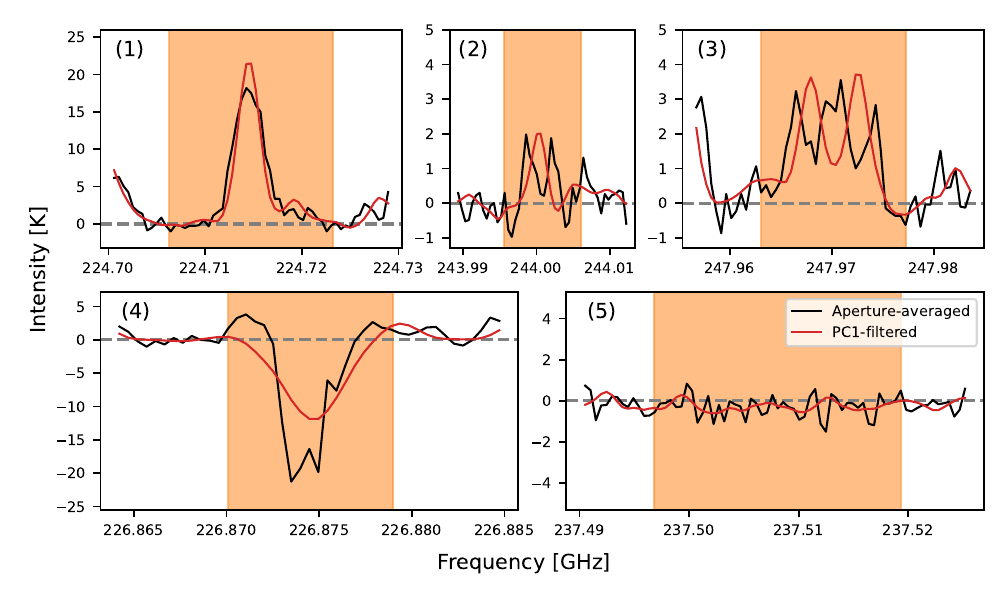}
\caption{The example of the selected frequency ranges and their PC1-filtered spectra. Each panel shows the aperture-averaged (black) and PC1-filtered spectra (red) within the selected frequency ranges (the orange filled regions); they are one of the (1) Gaussian-like emission lines with high peak intensities, (2) Gaussian-like lines but with low intensities, (3) blended broad emission lines, (4) deep absorption features, and (5) emission free-regions. As presented in the upper right panel, if two adjacent lines are intrinsically blended because of thermal and non-thermal effects or a low spectral resolution, those lines cannot be completely decoupled even with the PC1-filtering method.}
\label{fig_sel_freqs}
\end{figure*}

We also check whether the PC1-filtering method preserves intensity in general. In order to take account of various types of lines, we selected 109 frequency ranges which contain (1) Gaussian-like emission lines with high peak intensities, (2) Gaussian-like lines but with low intensities, (3) blended broad emission lines, (4) deep absorption features, and (5) emission free-regions. Figure \ref{fig_sel_freqs} shows examples of the selected frequency ranges. Among the 109 frequency ranges selected for this test, two frequency ranges are located within the second SPW as marked by the orange color shades in Figure \ref{fig_filter_spec}. 

The right panel of Figure \ref{fig_Int_conserv} shows that the integrated intensities from the PC1-filtered spectra are consistent with those from the aperture-averaged spectra for all selected lines. The rms of the intensity differences is 1.08~K~km~s$^{-1}$. Since the mean error of the integrated intensities from the PC1-filtered spectra is about 0.97~K~km~s$^{-1}$, the measured integrated intensities are consistent with each other within about 1-$\sigma$ error range. As seen in the bottom panel, the residuals are randomly distributed around zero. Thus, the slightly lower intensities of the PC1-filtered spectra above 40~K~km~s$^{-1}$ for the isolated strong lines may be a coincidence resulting from a lack of samples in the high intensities. Therefore, the PC1-filtering method preserves the total intensity of a line.

\subsection{Comparison with the other methods}

\begin{figure*}[htp]
\centering
\includegraphics[width=0.9\textwidth]{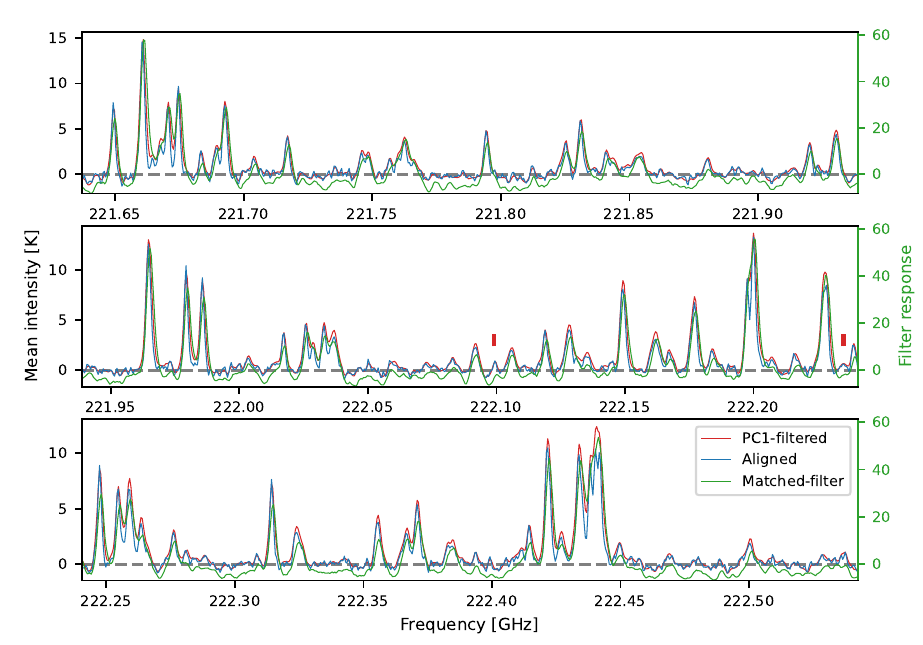}
\caption{Comparison of spectra produced by the PC1-filtering (red), aligned (blue), and matched filter methods (green). Note that the matched filter method provides the filter response values instead of the intensities. The red bars indicate two tentatively identified COM lines that are detected in the PC1-filtered spectrum over the 1-$\sigma$ level, which are not visible in the matched-filtered spectrum.}
\label{fig_comp_mf}
\end{figure*}

Figure \ref{fig_comp_mf} presents a comparison of the PC1-filtered spectrum (red), aligned spectrum (blue), and matched-filtered spectrum (green) obtained from the water-sublimated region of V883 Ori. There are slight variations among the three spectra in terms of relative line intensities, line profiles, and noise patterns near zero intensity. These discrepancies are likely attributed to the use of different emission structures ($T_\mathrm{PC1}$ for the PC1-filtering method, moment 1 map for the aligning method, and kernel for the matched-filter method), which may not perfactly coincide with each other. 

The aligned spectrum shows line intensities consistent with those in the PC1-filtered spectrum. However, the lines detected in the PC1-filtered spectrum tend to have slightly broader line widths (FWHM $\sim$ 3.28~km~s$^{-1}$) and smoother line profiles compared to those in the aligned spectrum (FWHM $\sim$ 2.40~km~s$^{-1}$). These differences are likely a result of the convolution of the cube data and window function, which has a similar effect to Gaussian filtering on the spectra. On the other hand, the aligned spectrum has the T$_\mathrm{rms}$ value of 0.88~K, which is comparable with that of the aperture-averaged spectum (0.81~K). As both the aperture-averaged and aligned spectra are extracted from similar apertures encompassing the water-sublimated region of V883 Ori, they include a similar number of pixels to derive their mean spectra, resulting in similar noise levels. Thus, the SNR of the PC1-filtered spectrum is higher compared to the aligned spectra by a factor of $\sim$ 2. 

The matched-filtered spectrum exhibits overall features consistent with the other spectra. Similar to the PC1-filtered spectra, the matched-filtered spectra also show broader line widths (FWHM $\sim$ 3.34~km~s$^{-1}$) compared to the aligned spectra. Moreover, both PC1-filtered and matched-filtered spectra demonstrate similar improvements in SNR. The SNR of the matched-filtered spectra is calculated using the 1-$\sigma$ noise level estimated from their emission-free channels. The maximum SNR values for the PC1-filtered and matched-filtered spectra are 37.13 and 40.90, respectively. Since the maximum SNR of the aperture-averaged spectrum is 18.47, both PC1 filtering and matched filtering methods achieved approximately a twofold improvement in SNR. However, the matched-filtered spectrum has the relatively unstable baseline, which poses challenges in identifying weak emission lines. For instance, the red bars in Figure \ref{fig_comp_mf} indicate weak emission lines that are visible in the PC1-filtered and aligned spectra but not in the matched-filtered spectrum. Through model fitting, these lines are tentatively identified as weak COM lines over the 1-$\sigma$ level: one is a CH$_2$CCH line at 222.099~GHz, and the other is a H$_2$C$^{13}$CO line at 222.235~GHz. 

Each of the methods presented here has advantages and limitations. The aligning method and matched filtering method offer the ability to obtain line spectra easily. In particular, the matched filtering method allows us to derive a high SNR spectrum from visibility data without requiring the CLEAN process. Therefore, a quick search for specific transition lines can be achieved with low computational costs. However, both methods require prior information on the kinematic structure of a target. In addition, despite a narrower line width, the aligning method is accompanied by a lower SNR than the other methods because the aligned spectrum includes all noise signals in the vicinity of the line emissions. The matched-filtered spectrum significantly improves SNR as the contribution of noise signals is efficiently reduced by the kernel. However, line identification using the matched-filtered spectrum via model fitting can be challenging due to an unstable baseline. The matched-filtered spectrum also does not provide information on the line intensities, so we need subsequent CLEAN processes.

On the other hand, for the PC1-filtering method, initial line identification is required to find the isolated strong lines. The statistical accuracy of the emission distribution in $T_\mathrm{PC1}$ drops if there are few isolated strong lines. The PC1-filtering method, however, derives a window function from multiple emission lines and obtains a representative emission distribution. Therefore, this method is very useful when we carry out unbiased spectral surveys, especially for the COM lines. We can acquire high-SNR spectra without prior information on the kinematic structure of a target. Despite the broader line width compared to the aligned spectra, the high SNR and a relatively stable baseline of the PC1-filtered spectra make them capable of detecting weak emission lines, which are crucial to studying COMs. Furthermore, unlike the matched-filtering method, the PC1-filtering method can provide accurate intensities of the observed lines.

\subsection{Application of the PC1-filtering method}

The PC1-filtering method produces a representative spectrum of a line emitting gas traced by PC1, which presents a common emission structure of these observed lines. This method has only two requirements: an image cube with many emission lines and the line catalog covering the observed frequencies. Even with no prior information on the kinematics of the target, the PCA can assess the velocity structures traced by the selected lines. These advantages make the PC1-filtering method robust to extract representative spectra of the target. Also, the PC1-filtering method can extract high SNR spectra with a single Gaussian profile from the observed cube data. Also, weak lines, which cannot be identified from the original cube data, will rise up above the now reduced noise level.

Recently, ALMA has been utilized for unbiased line surveys of COMs (\citealt{Jor16}; Lee et al. submitted). For these survey data, our PC1-filtering method can be used to identify COM lines, including weak lines. With those identified lines, the physical and chemical environments of the COM emission region will be investigated more accurately; the high SNR spectra with single Gaussian profiles provide good quality data sets to fit with line simulation tools, such as XCLASS \citep{Mol17} and the MAdrid Data CUBe Analysis \citep[MADCUBA; ][]{Mar19}. By the fitting process, the blended spectra can also be decomposed. 

\begin{figure}[htp]
\centering
\includegraphics[width=0.4\textwidth]{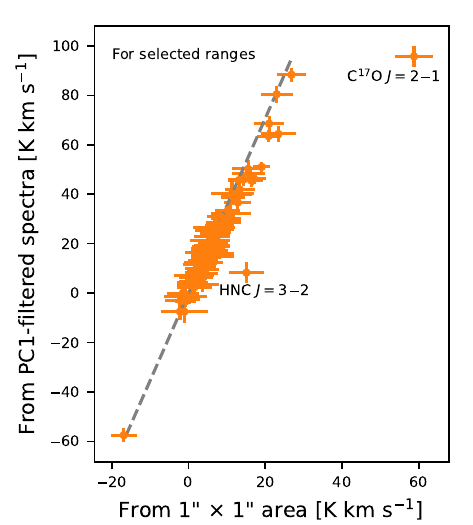}
\caption{The same as the right panel of Figure \ref{fig_Int_conserv} but for a comparison between the PC1-filtered spectra and the average spectra over the 1\arcsec\ by 1\arcsec\ area centered on V883 Ori. The gray dashed line shows $y=ax$, where $a$ is the area ratio between the circular aperture of r$=$0.3\arcsec\ and the boxy aperture of 1\arcsec\ by 1\arcsec.}
\label{fig_search_extend}
\end{figure}

In addition, we can utilize the PCA products to identify lines with emission distribution different from the COM emission. Figure \ref{fig_search_extend} shows the comparison of the integrated intensities between the PC1-filtered spectra and the spectra averaged over the 1\arcsec\ by 1\arcsec\ boxy aperture, where the PC1 has been derived, for the same selected frequency ranges as presented in Section 5.1. The data points follow a linear relation described by the area ratio of the circle with r$\sim$0.3\arcsec\ and the box of 1\arcsec\ by 1\arcsec\ (the gray dashed line). This result is expected because the integrated intensities of the PC1-filtered spectra agree with those of the spectra extracted from the circular aperture of r$\sim$0.3\arcsec\ (Figure \ref{fig_Int_conserv}). However, two outliers corresponding to C$^{17}$O $J=$2$-$1 and HNC $J=$3$-$2 appear. These two lines trace much more extended structures compared to the COM lines (\citealt{JELee19}; \citealt{Tob23}; Lee et al. submitted): the COM lines are confined within the water sublimation radius traced by the HDO emission, while the C$^{17}$O line traces the whole dust disk structure, and the HNC line traces a ring-like structure beyond the C$^{17}$O emission.

\begin{figure}[htp]
\centering
\includegraphics[width=0.4\textwidth]{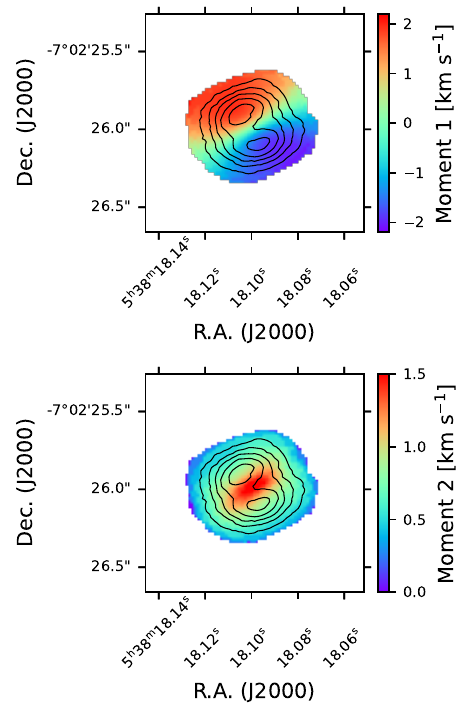}
\caption{Moment 1 (top) and moment 2 (bottom) maps for the eigencube of PC1. The black contours in both panels show the moment 0 map of PC1 (presented in the upper-left panel of Figure \ref{fig_PC1_Eigcube}.}
\label{fig_mom12}
\end{figure}

The $T_\mathrm{PC1}$ also provides information on the kinematics of the target with high SNR. As presented above, the COM lines trace the inner part of the Keplerian disk of V883 Ori. The channel maps in Figure \ref{fig_PC1_Eigcube} exhibit that the noise signals in $T_\mathrm{PC1}$ is much lower than those in the original cube data. Figure \ref{fig_PV} also shows that the maximum SNR of the PV-diagram of $T_\mathrm{PC1}$ ($\sim$ 197.6) is much higher than that of Line 1 ($\sim$ 26.4). This improved SNR of PC1 greatly enhances kinematics analysis capabilities. Figure \ref{fig_mom12} shows the maps of intensity-weighted velocity (moment 1) and intensity-weighted velocity dispersion (moment 2) for the eigencube of PC1. These maps are commonly used in radio astronomy to assess the kinematics of the target. The SNR of these maps depends on how many noise signals are included, so the moment maps have been produced using a mask that can exclude the noise signals \citep{Dam11}. In the PC1-filtering method, the randomly varying noise signals are suppressed in $T_\mathrm{PC1}$ because it describes the common feature of selected lines. Therefore, the moment maps derived from $T_\mathrm{PC1}$ have very high SNR.

The $T_\mathrm{PC1}$ traces the inner disk of V883 Ori inside the water sublimation radius because most COM lines trace that component. Lee et al. (submitted) categorize observed molecular lines into several groups which trace different spatial and kinematic structures of V883 Ori. Therefore, we can construct additional versions of $T_\mathrm{PC1}$ that trace these kinematics using other molecular emission lines. If a molecular line traces a mixture of different kinematics, including the inner rotating disk, we could decompose these different kinematics. First, the emission components tracing the inner disk can be inferred by fitting the emission structure with $T_\mathrm{PC1}$. After removing the inner disk component from the cube data, we can explore other kinematic features step by step using the filters derived from the PCA.

\section{Summary} \label{sec_sum}
We introduce the PC1-filtering method, which is robust to extract the representative spectra of a line-rich target. This method derives a common emission structure of selected emission lines in position-position-velocity space using PCA and utilizes the structure to obtain a representative spectrum of the gas component traced by the selected lines. Thus, this method provides a high-SNR kinematics-corrected line spectrum without requiring any prior information about the target. We apply the PC1-filtering method to the ALMA Spectral Survey of An eruptive Young star, V883 Ori (ASSAY) in Band 6 (2019.1.00377.S, PI: Jeong-Eun Lee), especially for the COM lines. The main results are summarised as follows: 

\begin{enumerate}
\item We find 34 COM lines that are strong and well-isolated from adjacent lines. These lines are confined within the water sublimation radius (within 0.3 \arcsec\ from the center), positively correlated, and emitted from the same gas component. The PC1 of PCA-3D exhibits their common emission structure, and the emission structure follows the kinematics of the Keplerian rotating disk of V883 Ori.
\item The PC1-filtered spectra are derived by utilizing $T_\mathrm{PC1}$. The PC1-filtering method corrects non-Gaussian line profiles generated by the rotating disk and produces line spectra with single Gaussian line profiles with higher peak intensities. 
\item The PC1-filtering method effectively decreases T$_\mathrm{rms}$ in the spectra by a factor of 2 compared to the aperture-averaged spectra directly extracted from the image cube. This reduction in noise, along with the corresponding enhancement in line peak intensity, results in an overall improvement of the SNR by a factor of 2.5. This level of improvement is comparable to what can be achieved using the matched-filtering method.
\item The PC1-filtered spectra are generally consistent with the filter response spectra obtained using the matched filter method. However, we have observed certain cases where the PC1-filtered spectra reveal weak emission lines that are not easily identified in the matched-filtered spectra. This finding highlights the usefulness of the PC1-filtered spectra in detecting and identifying weak emission lines that might otherwise be missed using the matched-filtering method.
\item The PC1-filtering method preserves the integrated intensities of the observed lines. The integrated intensities for emission lines measured from the PC1-filtered spectra are consistent with those from the aperture-averaged spectra using the 3-$\sigma$ criterion. 
\item By comparing the integrated intensities of the PC1-filtered spectra and those from the spectra averaged over the 1\arcsec\ by 1\arcsec\ boxy aperture, we can identify the specific lines tracing different gas components.
\item The PC1-filtering method can be applied to any unbiased spectral survey data set to extract kinemics-corrected spectra. Thus, it can be very useful, especially, for the COM lines that are easily blended with each other. High-SNR kinematics-corrected line profiles are crucial when they are fitted with line simulation tools, such as XCLASS and MADCUBA, for further quantitative analysis. 
\item $T_\mathrm{PC}$ can be used to explore the system kinematics, with a high-SNR for the gas component of interest. We can also assess different kinematic features associated with the target by fitting out the emission structure with the primary $T_\mathrm{PC}$.
\end{enumerate}

\section{Acknowledgements}

We truly appreciate Doug Johnstone checking carefully the manuscript. This work was supported by the National Research Foundation of Korea (NRF) grant funded by the Korea government (MSIT) (grant number 2021R1A2C1011718).
This paper makes use of the following ALMA data: ADS/JAO.ALMA\#2019.1.00386T.
ALMA is a partnership of ESO (representing its member states), NSF (USA) and NINS (Japan), together with NRC (Canada), NSC and ASIAA (Taiwan), and KASI (Republic of Korea), in cooperation with the Republic of Chile. The Joint ALMA Observatory is operated by ESO, AUI/NRAO and NAOJ.

\clearpage
\appendix
\renewcommand\thefigure{\thesection.\arabic{figure}}
\renewcommand\thetable{\thesection.\arabic{table}}
\twocolumngrid
\section{Selected isolated strong lines} \label{App_isol_str} 
\setcounter{figure}{0}
\setcounter{table}{0}
We find 34 COM lines which are strong enough and well isolated from adjacent emission lines (see Section \ref{Sec_line_sel}). Table \ref{tbl_isol_str} shows the list of the selected \emph{isolated strong lines}, and Figure \ref{fig_IS_lines_mom0} shows their moment 0 maps.

\begin{deluxetable*}{llccccc}
\tablecolumns{7}
\tabletypesize{\scriptsize}
\tablecaption{List of the Isolated strong lines \label{tbl_isol_str}}
\tablewidth{0pt}
\tablehead{
          \colhead{Number\tablenotemark{a}}&\colhead{Molecule}&\colhead{Frequency}&\colhead{$I_\mathrm{aligned}$}&\colhead{E$_\mathrm{up}$}&\colhead{g$_\mathrm{up}$}&\colhead{A$_\mathrm{ij}$}\\
          \colhead{}&\colhead{}&\colhead{[GHz]}&\colhead{[K]}&\colhead{[K]}&\colhead{}&\colhead{}
          }
\startdata
1&      CH$_3$OH v=0&           261.80571&     11.28&  28&     20&     5.572$\times$10$^{-5}$\\
2&      CH$_3$OCHO v=0&         240.02114&     10.27&  122.3&  78&     2.006$\times$10$^{-4}$\\
3&      CH$_3$CHO v=0&          250.82916&     10.10&  120.3&  54&     5.064$\times$10$^{-4}$\\
4&      CH$_3$OH v=0&           234.68339&     16.04&  60.9&   36&     1.874$\times$10$^{-5}$\\
5&      CH$_3$CHO v=0&          270.41585&      8.93&  117.5&  58&     6.618$\times$10$^{-4}$\\
6&      CH$_3$OH v=0&           247.61096&     10.59&  446.6&  148&    8.290$\times$10$^{-5}$\\
7&      CH$_3$CHO v=0&          242.11814&     11.74&  83.8&   54&     4.996$\times$10$^{-4}$\\
8&      H$_2$CCO v=0&           260.19198&      9.33&  100.5&  81&     1.985$\times$10$^{-4}$\\
9&      CH$_3$OCHO v=0&         246.89161&      8.53&  126.2&  78&     2.179$\times$10$^{-4}$\\
10&     CH$_3$OCHO v=0&         260.24450&      7.75&  146.8&  86&     2.560$\times$10$^{-4}$\\
11&     CH$_3$OCHO v=0&         247.68266&      8.22&  166.7&  82&     1.930$\times$10$^{-4}$\\
12&     CH$_3$CHO v$_{15}$=1&   271.76847&      6.10&  322&    58&     6.799$\times$10$^{-4}$\\
13&     CH$_3$CHO v=0&          234.82587&     12.78&  81.8&   50&     4.449$\times$10$^{-4}$\\
14&     CH$_3$CHO v$_{15}$=1&   254.48892&      7.38&  294.5&  54&     5.852$\times$10$^{-4}$\\
15&     CH$_3$OCHO v=0&         255.78941&      7.74&  147.3&  86&     2.420$\times$10$^{-4}$\\
16&     CH$_3$OCHO v=0&         248.63369&      7.78&  157&    82&     1.964$\times$10$^{-4}$\\
17&     CH$_3$OCHO v=0&         237.82983&     11.06&  136.8&  78&     1.829$\times$10$^{-4}$\\
18&     CH$_3$CHO v$_{15}$=1&   271.35554&      6.34&  310.3&  58&     6.886$\times$10$^{-4}$\\
19&     CH$_3$OCHO v=0&         225.60882&     11.52&  116.7&  78&     1.669$\times$10$^{-4}$\\
20&     CH$_3$CHO v$_{15}$=1&   252.68125&      6.89&  299.8&  54&     5.541$\times$10$^{-4}$\\
21&     CH$_3$CHO v$_{15}$=1&   252.14430&      6.37&  309&    54&     5.374$\times$10$^{-4}$\\
22&     CH$_3$CHO v=0&          242.01030&      7.80&  104.6&  54&     4.785$\times$10$^{-5}$\\
23&     CH$_3$OH v$_{12}$=1&    247.84013&      7.28&  545.1&  100&    6.281$\times$10$^{-5}$\\
24&     CH$_3$CHO v=0&          238.67703&      9.89&  92.5&   50&     3.547$\times$10$^{-5}$\\
25&     CH$_3$CHO v=0&          247.43300&      5.97&  28.5&   26&     2.578$\times$10$^{-5}$\\
26&     CH$_3$CHO v$_{15}$=1&   233.04852&      9.43&  285.1&  50&     4.314$\times$10$^{-4}$\\
27&     CH$_3$OCHO v=0&         237.80763&      9.26&  136.8&  78&     1.825$\times$10$^{-4}$\\
28&     CH$_3$OCHO v=0&         235.84454&      9.78&  145&    78&     1.713$\times$10$^{-4}$\\
29&     H$_2$CCO v=0&           222.31440&      9.65&  116.2&  23&     1.196$\times$10$^{-4}$\\
30&     c-C$_2$H$_4$O v=0&      225.46834&      8.84&  32.5&   55&     1.058$\times$10$^{-4}$\\
31&     CH$_3$CHO v=0&          240.75590&      5.56&  92.6&   50&     1.027$\times$10$^{-5}$\\
32&     c-C$_2$H$_4$O v=0&      226.04315&      8.34&  47&     45&     1.661$\times$10$^{-4}$\\
33&     CH$_3$OCHO v=0&         245.67298&      5.58&  273.1&  82&     9.836$\times$10$^{-5}$\\
34&     H$_2$CCO v=0&           242.42466&      6.10&  127.8&  25&     1.565$\times$10$^{-4}$\\
\enddata
\tablenotetext{a}{The selected lines are arranged in a decreasing order of the correlation coefficient with PC1 of PCA-3D (see Section \ref{sec_rst_PCA}).}
\end{deluxetable*}

\begin{figure*}[htp]
\centering
\includegraphics[width=1.0\textwidth]{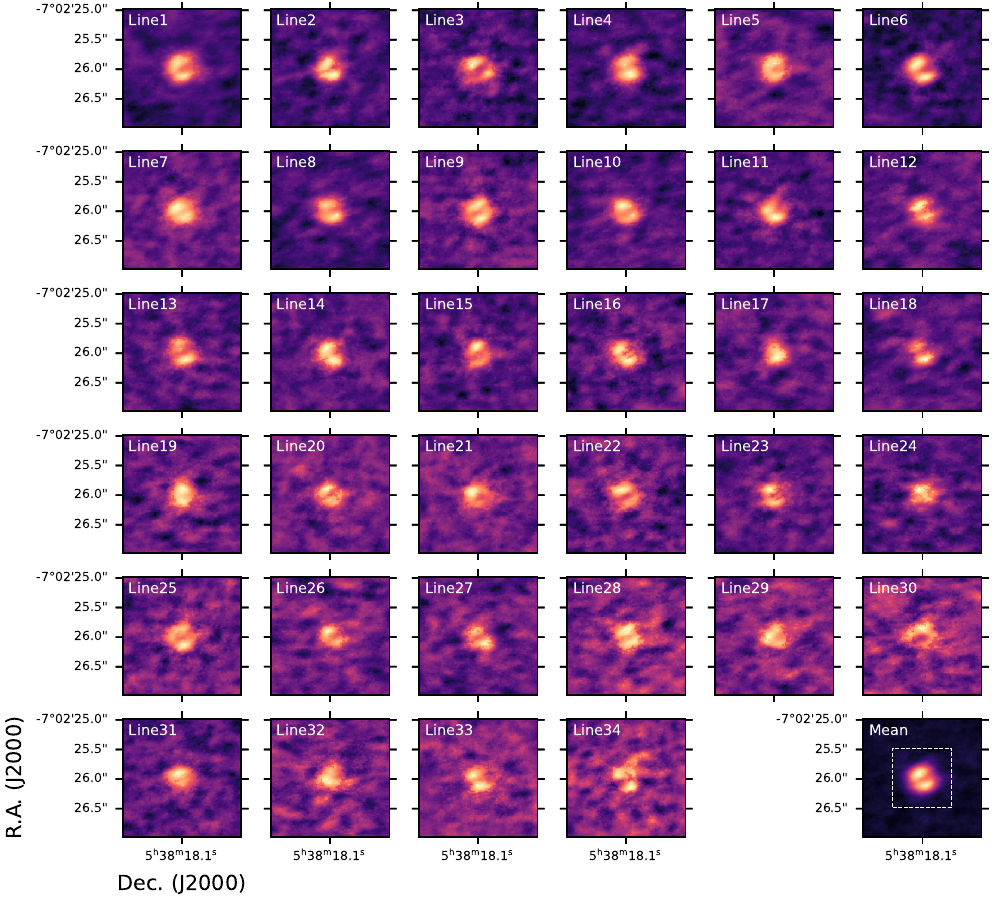}
\caption{Moment 0 maps for the 34 isolated strong lines. The lines are named by the numbers listed in Table~\ref{tbl_isol_str}. The bottom-right panel presents a mean moment 0 map of the 34 lines. the white dashed box in the mean moment 0 map represents 1\arcsec\ by 1\arcsec\ area centered on V883 Ori.}
\label{fig_IS_lines_mom0}
\end{figure*}

\section{Position-velocity diagram of V883 Ori} \label{App_PV} 
\setcounter{figure}{0}
\setcounter{table}{0}

\begin{figure*}[htp]
\centering
\includegraphics[width=0.9\textwidth]{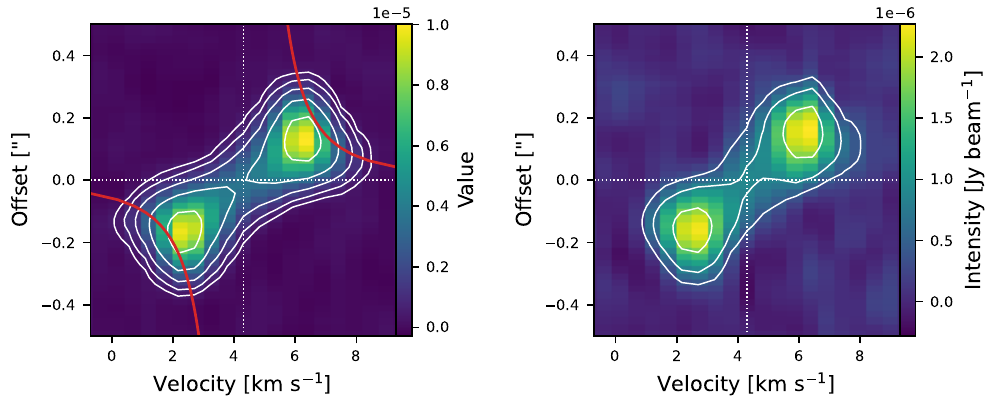}
\caption{The position-velocity (PV) diagrams extracted along the semi-major axis of the V883 Ori disk (the magenta dashed line in Figure \ref{fig_PC1_Eigcube}). The left panel shows the PV diagram for $T_\mathrm{PC1}$, and the right panel shows that for the Line 1. In each panel, the contour levels starts from 5 times the noise level of the data and are spaced by a factor of 2. The red solid lines on the left panel depict the Keplerian rotation profile of the V883 Ori disk.}
\label{fig_PV}
\end{figure*}

The position-velocity (PV) diagram is a commonly used analysis to assess the kinematics of a rotating disk. We generate the PV diagram of V883 Ori along the semi-major axis of its disk. Figure \ref{fig_PV} shows the PV diagram of $T_\mathrm{PC1}$. The distribution of the COM lines can be explained by a Keplerian rotation profile (the red solid line) around a 1.2~M$_\sun$ central protostar \citep{Cie16, JELee19}. The inner boundary of the emission ($\sim$0.1\arcsec) is set by the optically thick continuum emission, while the outer boundary ($\sim$0.3\arcsec) is determined by the water sublimation radius.

\clearpage
\bibliographystyle{aasjournal}
\bibliography{ref}{}

\end{document}